\documentstyle[prd,aps,preprint,eqsecnum]{revtex}
\begin{document}
\draft
\newcommand{\barbox}{\stackrel{-}{\Box}}
\newcommand{\chp}{\rm{ch}\hspace{1pt}{\it p}}
\newcommand{\shp}{\rm{sh}\hspace{1pt}{\it p}}
\title{A dynamical symmetry breaking model in Weyl space}
\author{A. Feoli}
\address{Dipartimento di Scienze Fisiche ``E.R.Caianiello''
 Universit\`{a} di Salerno
I- 84081 Baronissi (Salerno), Italy}
\author{W.R. Wood}
\address{Faculty of Natural and Applied Sciences,
Trinity Western University, 7600 Glover Road,
Langley, British Columbia, V2Y 1Y1, Canada}
\author{G. Papini}
\address{Department of Physics, University of Regina,
Regina, Saskatchewan, S4S 0A2, Canada}
\maketitle

\begin{abstract}
The dynamical process following the breaking of Weyl geometry to
Riemannian geometry is considered by studying the motion of de Sitter
bubbles in a Weyl vacuum.  The bubbles are given in terms of an exact,
spherically symmetric thin shell solution to the Einstein equations in a
Weyl-Dirac theory with a time-dependent scalar field of the form $\beta =
f(t)/r$.  The dynamical solutions obtained lead to a number of possible
applications.  An important feature of the thin shell model is the manner in
which $\beta $ provides a connection between the interior and exterior geometries since information about the exterior geometry is contained in
the boundary conditions for $\beta$. 
\end{abstract}


\section{Introduction}
The success of models like the Glashow-Weinberg-Salam electroweak model
have provided strong motivation for believing that symmetries of nature
that were once manifest are no longer directly observable; one must
extrapolate backwards from our broken symmetry state to the symmetric state
that is described by the Lagrangian.  The ultimate success of the standard
model is contingent upon the existence of the Higgs particle---unless an
additional degree of symmetry is broken \cite{Che}.  This latter
possibility, that involves the breaking of Weyl's scale invariance, has
recently received further consideration \cite{Paw}.  While the topic
of breaking Weyl geometry to Riemannian geometry has a relatively
long history (see, e.g., \cite{Dom}), the analysis of the dynamics of thin shells that form in this process has been neglected.  This problem is considered here by studying the dynamical properties of spherically symmetric thin-shell
solutions to Einstein's equations with Weyl's scale invariance broken in
the interior of the bubble \cite{Woo92a}.

Shortly after Einstein presented his general theory of relativity, Weyl
\cite{Wey} attempted to unify the electromagnetic and gravitational fields
within a geometrical framework.  Einstein had shown that the rotation of a
vector under parallel transport was related to the gravitational field;
Weyl argued that the electromagnetic field would acquire a similar
geometric status if the length of the vector was also allowed to vary
according to
$\delta \ell =\ell \kappa _{\mu }\delta x^{\mu }$.  The covector field
$\kappa _{\mu }$, together with the metric tensor $g_{\mu \nu}$ that is
defined modulo an equivalence class, comprised the fundamental fields of
the new geometry.  Weyl's geometry was rejected at the time because it
didn't permit an atomic standard of length.  In more recent years, Weyl's
theory has been reconsidered from several points of view (see, e.g.\
\cite{Woo92a,Dir,Gre,San,Woo92b,Woo93a,Woo93b}).  Of particular interest
here is the suggestion that atomic standards of length are introduced into
a pure Weyl geometry through a symmetry breaking process.

The possibility that regions of spacetime with broken scale invariance can
coexist within regions of unbroken scale invariance has been demonstrated
\cite{Woo92a} by applying the Gauss-Mainardi-Codazzi (GMC) formalism in
Weyl geometry to a Weyl-Dirac theory with a real scalar field $\beta $.
The surface tension of the resulting static, spherically symmetric bubble
arose from the boundary conditions imposed on the normal component of the
derivative of the field $\beta $.  In this way, a stable ``particle'' was
constructed entirely from the fields $g_{\mu \nu }$, $\kappa _{\mu }$, and
$\beta $.  By setting $\kappa _{\mu }=0$ and $\beta = $ constant in the
interior space, the conformal invariance is broken and an ``atomic'' scale
can be introduced into the theory.  Generalizations of this model have been
considered \cite{Woo92b,Woo93a,Woo93b}, including the development of a
geometric formulation of the causal interpretation of quantum mechanics
\cite{Woo93a} where $\beta $ is a complex scalar field.  In the present
work, $\beta $ is taken to be real and of the form $\beta (r,t) = f(t)/r$
in the exterior space $V^{E}$.  This choice is useful in focussing on the
dynamical properties of bubbles after they emerge in the ambient Weyl
geometry.

The absence of a generalized Birkhoff theorem in Weyl space allows
dynamical solutions to exist in the spherically symmetric case in the model
under consideration.  The
analysis of exact, time-dependent solutions to the radial equation of
motion  shows that spherically symmetric bubbles that form in
the ambient Weyl vacuum may, depending on the circumstances, collapse,
remain at a fixed radius, oscillate indefinitely, or diverge to infinity.
The spherically symmetric solution to the Einstein equations for the
ansatz $\beta (r,t) = f(t)/r$ is presented in Sec.\ 2.  The analysis of the
radial motion of the thin shell is provided in Sec.\ 3 and some concluding
remarks are given in Sec.\ 4.

\section{The spherically symmetric solution}

The GMC formalism is here applied to the Weyl-Dirac theory given by
\begin{eqnarray}
I_{D}=\int \{-\frac{1}{4}f_{\mu \nu }f^{\mu \nu }+
\beta ^{2}R+6\beta _{,\mu }\beta ^{,\mu }-
\lambda \beta ^{4}\}\sqrt{-g}d^{4}x, \label{2.1}
\end{eqnarray}
where $f_{\mu \nu }=\kappa _{\nu ,\mu }-\kappa _{\mu ,\nu }$.
This action was introduced by Dirac \cite{Dir} in order to study his 
Large Numbers hypothesis within the context of a theory of gravitation.  
Dirac's eloquent reasoning led him to propose a two-metric view that
served to revive interest in Weyl's geometry.  Generalizations of Dirac's
action were subsequently proposed (see, e.g., \cite{Gre}) which accommodated
Weyl's geometric interpretation of the electromagnetic field within a single-metric
theory.  This latter view is also maintained in the present paper.

The field equations that follow from (\ref{2.1})
are the Maxwell and Einstein equations
\begin{eqnarray}
\Box _{\nu }f^{\mu \nu }=0, \label{2.2}
\end{eqnarray}
and
\begin{eqnarray}
G_{\mu \nu }=\frac{1}{2\beta ^{2}}E_{\mu \nu }+
I_{\mu \nu }+\frac{1}{2}\lambda g_{\mu \nu }\beta ^{2}
\equiv T_{\mu \nu }, \label{2.3}
\end{eqnarray}
where $E_{\mu \nu }$ is the usual Maxwell tensor and $\Box $ is the
Riemannian covariant derivative,
\begin{eqnarray}
I_{\mu \nu }=\frac{2}{\beta}(\Box _{\nu }\Box _{\mu }\beta -
g_{\mu \nu }\Box _{\alpha }\Box ^{\alpha }\beta )-
\frac{1}{\beta ^{2}}(4\beta _{,\mu }\beta _{,\nu }-
g_{\mu \nu }\beta _{,\alpha }\beta ^{,\alpha }), \label{2.4}
\end{eqnarray}
and the field equation for $\beta $ is identically the trace
of (\ref{2.3}).

For simplicity, the GMC formalism is applied to a spherically symmetric
shell.   The most general line element is then given by \cite{Syn}
\begin{eqnarray}
ds^{2}=-e^{\nu (r,t)}dt^{2}+e^{\mu (r,t)}dr^{2}+
r^{2}(d\theta ^{2}+\sin ^{2}\theta d\phi ^{2}). \label{2.5}
\end{eqnarray}
The exterior and interior geometries are distinguished by
writing $t_{E,I}$, $\nu _{E,I}$ and $\mu _{E,I}$ in the exterior and interior
spacetimes $V^{E,I}$, respectively.  It will be assumed here that the
bubbles which form in the symmetry breaking process have interior
geometries
with $\kappa _{\mu }=0$ (which establishes length integrability
in the region of spacetime occupied by the bubble) and $\beta =\beta _{0}$
(which breaks the conformal invariance and fixes the scale in $V^{I}$
according to the value of the constant
$\beta _{0}$).  Due to their continuity properties,
$\kappa _{\mu }=0$ and $\beta =\beta _{0}$ on the thin shell
$\Sigma $ as well \cite{Woo92a}.
With the Maxwell tensor and $I_{\mu \nu }$ vanishing in $V^{I}$,
the interior stress-energy tensor reduces to
\begin{eqnarray}
T^{I}_{\mu \nu }=\frac{1}{2}\lambda g_{\mu \nu }\beta _{0}^{2}.
\label{2.6}
\end{eqnarray}
In what follows, $\lambda $ will be taken to be negative.  For this choice,
and assuming no additional matter sources in $V^{I}$,
the tensor (\ref{2.6}) represents
a de Sitter space and the interior metric is given by
\begin{eqnarray}
e^{-\mu _{I}}=1+\frac{1}{6}\lambda \beta _{0}^{2}r^{2}=
e^{\nu _{I}}, \label{2.7}
\end{eqnarray}
with a horizon at $ r = \beta_{0}^{-1} \sqrt{6/|\lambda|}
 \equiv r_{h}^{I}$.

In the exterior geometry $V^{E}$, $\beta $ is expressed in the form
\begin{eqnarray}
\beta (r,t) = f(t)/r, \label{2.8}
\end{eqnarray}
where $f(t)$ is an arbitrary dimensionless function of $t$, and where
$t\equiv t_{E}$ here and henceforth.  Normally, one would proceed in
applying the GMC formalism by solving the Einstein equations in $V^{E}$ for
the ansatz (\ref{2.8}) and then using the resulting solution to obtain the
equation of motion for the thin shell.  In the present model  a number of
possible
exterior metrics are known to exist.  However, the condition that
$\beta = \beta _{0}$ in $V^{I}$ together with the boundary condition that
$\beta $ be continuous across $\Sigma $ for all $t$ requires that $\beta $
also be a constant with respect to the intrinsic time $\tau $ of the thin
shell defined at $r = R(\tau )$.  That is,
\begin{eqnarray}
\frac{d\beta }{d\tau }|_{r=R} = \beta _{,t}X_{E} + \beta ' \dot{R} = 0,
\label{2.9}
\end{eqnarray}
where $X_{E}\equiv dt/d\tau $ and the prime and dot denote differentiation
with respect to $r$ and $\tau $, respectively.  The resulting condition,
\begin{eqnarray}
X_{E} = - \frac{\beta '}{\beta _{,t}} \dot{R}, \label{2.10}
\end{eqnarray}
leads to a restriction on the set of possible solutions to (\ref{2.3}) in
$V^{E}$. In effect, the condition (\ref{2.10}) establishes a relationship
between the radial and time derivatives of $\beta $ at $\Sigma $ that
constrains the exterior metric since it is a functional of $\beta $.  This
constraint is satisfied when the exterior metric takes the form (see the
Appendix)
\begin{eqnarray}
e^{\nu _{E}}=-\frac{\beta _{,t}^{2}}{\beta '^{2}(1+
\frac{1}{6}\lambda \beta ^{2}r^{2})} \label{2.11}
\end{eqnarray}
and
\begin{eqnarray}
e^{\mu _{E}} = -\frac{1+\frac{1}{6}\lambda \beta ^{2}r^{2}}{\gamma ^{2}
\beta ^{2}r^{4}}, \label{2.12}
\end{eqnarray}
where $\gamma $ is an arbitrary constant with dimension (length)$^{-1}$.
If one assumes that the sign of the constant $\lambda $ does not change
during the formation of the bubble, then it must be taken to be negative to
ensure the correct signature of the metric (see \cite{Woo92a} for the
conventions used here).  For this choice, the signature $(-,+,+,+)$ is
preserved for $r^{2}<(r_{h}^{I})^{2}$ in $V^{I}$ and for
$r^{2}>(r_{h}^{I})^{2}\beta _{0}^{2}/\beta ^{2}$ in $V^{E}$.
For the ansatz (\ref{2.8}) this latter condition
becomes $f^2(t) > (r_{h}^{I})^{2}\beta _{0}^{2}$, so that the possibility of
a time-dependent change of signature exists.

It is interesting to note that, in the spherically symmetric case with
$\beta $ of the form given in (\ref{2.8}), the Einstein equations
(\ref{2.3}) also support the time-dependent solution
\begin{eqnarray}
e^{\nu _{E}}=-\frac{\beta _{,t}^{2}}{\beta '^{2}(1+
\frac{1}{6}\lambda \beta ^{2}r^{2}+
\frac{q^{2}}{4\beta ^{2}r^{2}})} \label{2.13}
\end{eqnarray}
and
\begin{eqnarray}
e^{\mu _{E}} = -\frac{1+\frac{1}{6}\lambda \beta ^{2}r^{2}+
\frac{q^{2}}{4\beta ^{2}r^{2}}}{\gamma ^{2}
\beta ^{2}r^{4}}. \label{2.14}
\end{eqnarray}
However, when the exterior metric is matched onto the thin shell via the
field $\beta $, the boundary conditions lead to the requirement that $q=0$.
This suggests that, in the time-dependent, spherically symmetric case, a
charged particle solution is not allowed.  In contrast, the surface charge
was an integral component of the static solution presented in
\cite{Woo92a}.

\section{The motion of the thin shell}

With the interior and exterior metrics in hand, it is now possible to study
the motion of the thin shell.  In the GMC formalism, the equation of motion
follows from the jump in the $\theta \theta $-component of the extrinsic
curvature $K_{\mu \nu }$
across $\Sigma $.  In terms of the interior metric (\ref{2.7})
together with the ansatz (\ref{2.8}), this
equation can be expressed in the form (see (\ref{A.12}))
\begin{eqnarray}
1+\dot{R}^{2}+\frac{1}{6}\lambda \beta _{0}^{2}R^{2} =
e^{(\mu - \nu )_{E}}\frac{\dot{R}^{4}}{X_{E}^{2}}. \label{3.1}
\end{eqnarray}
Substituting (\ref{A.8}) for $X_{E}^{2}$ into (\ref{3.1}) yields
\begin{eqnarray}
1+\dot{R}^{2}+\frac{1}{6}\lambda \beta _{0}^{2}R^{2} =
\frac{\dot{R}^{4}}{\dot{R}^{2}+e^{-\mu _{E}}}. \label{3.2}
\end{eqnarray}
Using (\ref{2.12}) at $r=R$ (with $\beta = \beta _{0}$) and solving for
$\dot{R}^{2}$, one finds
\begin{eqnarray}
\dot{R}^{2} = \frac{a^{2}\left( 1-\frac{R^{2}}{R_{eq}^{2}}\right) }
{\left( \frac{R_{eq}^{2}}{R^{2}}-1\right) ^{2} - a^{2}} , \label{3.3}
\end{eqnarray}
where $a^{2} = \gamma ^{2}\beta _{0}^{2}R_{eq}^{4}$ and
\begin{eqnarray}
R_{eq} = \frac{1}{\beta _{0}}\sqrt{\frac{6}{|\lambda |}} = r_{h}^{I} \label{3.4}
\end{eqnarray}
results when $\dot{R} = 0$.  It is possible to obtain an analytical
description of the radial motion from (\ref{3.3}).  The analysis of the
motion of the thin shell is facilitated by breaking the problem into the
following two cases.

\subsection{Case 1:  $R<R_{eq}$}

The situation considered here applies to the case where a bubble of
Riemannian geometry with $R<R_{eq}$ forms in the ambient Weyl space.
In this case, the substitution of variables given by
\begin{eqnarray}
x\equiv \frac{R_{eq}^{2}}{R^{2}}-1 \label{3.5}
\end{eqnarray}
allows (\ref{3.3}) to be transformed into the integral form
\begin{eqnarray}
I_{1}\equiv \int \frac{1}{1+x}\sqrt{\frac{x^{2}-a^{2}}{x}}dx =
\pm 2\gamma \beta _{0}R_{eq}\tau . \label{3.6}
\end{eqnarray}
Setting $x=\cosh p\equiv \chp $, the integral $I_{1}$ becomes
\begin{eqnarray}
I_{1} = \frac{1}{\sqrt{a}}\left[
\int \frac{(a \chp -1)}{\sqrt{\chp }}dp +
(1-a^{2})\left( \int \frac{dp}{\sqrt{\chp }} -
a\int \frac{\sqrt{\chp }}{a\chp +1}dp
\right) \right] .
\label{3.7}
\end{eqnarray}
Expressing the three integrals in (\ref{3.7}) in terms of elliptic
integrals (see eqs.\ (2.464.9), (2.464.10) and (2.464.15) in \cite{Gra}),
one has
\begin{eqnarray}
I_{1} = \sqrt{2a}\left\{ \sqrt{2}\frac{\shp }{\sqrt{\chp }} -
2E\left( \alpha , \frac{1}{\sqrt{2}}\right) +
(1-a)\left[ F\left( \alpha , \frac{1}{\sqrt{2}}\right) -
\Pi \left( \alpha ,\frac{1}{a+1},\frac{1}{\sqrt{2}}\right)
\right] \right\} , \label{3.8}
\end{eqnarray}
where $p \neq 0$, $\shp \equiv \sinh p$ and
\begin{eqnarray}
\alpha \equiv \arcsin \sqrt{\frac{\chp -1}{\chp }}=
\arcsin \sqrt{\frac{R_{eq}^{2}-2R^{2}}{R_{eq}^{2}-R^{2}}} .  \label{3.9}
\end{eqnarray}

The case $a=1$ can be investigated easily because it involves only the
elliptic integral of the second kind $E(\alpha , 1/\sqrt{2})$.  This choice
does not represent a serious limitation due to the arbitrariness of the
parameters $\gamma $ and $\beta _{0}$ that appear in $a$.  With this
choice, Eq.\ (\ref{3.6}) becomes
\begin{eqnarray}
-\sqrt{2}E\left( \alpha , \frac{1}{\sqrt{2}}\right) +
\frac{\shp }{\sqrt{\chp }} = \pm \frac{\tau }{R_{eq}}, \label{3.9a}
\end{eqnarray}
which is valid for $0<R<R_{eq}/\sqrt{2}$.  For $R=0$ ($\alpha = \pi /2$,
$\chp = \infty $) the second term in (\ref{3.9a}) diverges, while
$R=R_{eq}/\sqrt{2}$ leads to $\alpha = 0$ and $\chp =1$ that invalidate
the integrations leading to (\ref{3.8}).  While $R$ cannot be set equal
to the endpoints of the interval over which it is defined (for $a=1$), it
is of interest
to consider the properties of the thin shell in the neighborhood of these
points by means of expansions.  To this end, it is useful to first invert
(\ref{3.9}), so that
\begin{eqnarray}
\frac{R}{R_{eq}} = \sqrt{\frac{1-\sin ^{2}\alpha }{2-\sin ^{2}\alpha }},
\label{3.9b}
\end{eqnarray}
and to choose units for which $R_{eq}=1$.


\subsubsection{Upper limit:  $R\approx R_{eq}/2^{1/2}$}

This limit corresponds to $\alpha \approx 0$.  The expansion of
(\ref{3.9a}) to third order in $\alpha $ yields
\begin{eqnarray}
\pm \frac{\tau }{2}-\frac{\alpha ^{3}}{3}\left(
\frac{1}{4}+\frac{1}{2^{3/2}}\right) =0 \label{3.9c}
\end{eqnarray}
which has the solutions $\alpha = \pm 1.52042 \hspace{1pt}\tau ^{1/3}$.
Substituting this result into (\ref{3.9b}), one obtains an oscillatory
motion with a period that increases with $\tau $ and
that has a maximum amplitude $R \sim R_{eq}/\sqrt{2}$.


\subsubsection{Lower limit:  $R\approx 0$}

This limit corresponds to $\alpha \approx \pi /2$.  Expanding (\ref{3.9a})
to second order in $\alpha - \pi /2$ for the $+$ sign on the
right-hand-side, one
obtains
\begin{eqnarray}
-\frac{1}{\alpha - \pi /2}+\tau E\left( \frac{1}{\sqrt{2}} \right)
+\left( \sqrt{1-\frac{1}{\sqrt{2}}} - \frac{1}{3(2)^{3/2}} \right)
\left( \alpha - \frac{\pi }{2}\right ) = 0, \label{3.9d}
\end{eqnarray}
where $E(1/\sqrt{2})$ is the complete elliptic integral of the second type.
The two solutions of (\ref{3.9d}), which are given by
\begin{eqnarray*}
\alpha = 0.109314 - 0.835143 \hspace{1pt}\tau \pm 0.835143 \sqrt{5.45723 +
3.4996 \hspace{1pt}\tau + \tau ^{2}},
\end{eqnarray*}
also lead to oscillatory motion when they are substituted into
(\ref{3.9b}).

The expansion for the $-$ sign on the right-hand-side of (\ref{3.9a})
leads to the two solutions
\begin{eqnarray*}
\alpha = 0.109314 + 0.835143 \hspace{1pt}\tau \pm 1.67249 \sqrt{1.36071 -
0.872685 \hspace{1pt}\tau + 0.249342 \hspace{1pt}\tau ^{2}}.
\end{eqnarray*}
Of these, the one with the $+$ sign in front of the square root yields
oscillatory motion.  The other solution, on the contrary, expands from
$R \sim 0.4 R_{eq}$ at $\tau = 0$ to the maximum $R \sim R_{eq}$
at $\tau \sim 1.5 $, and then decreases asymptotically to $R = 0$.

In summary, it is has been shown that bubbles which are
created in the Weyl vacuum with initial radii near either of the endpoints
of the interval $0<R<R_{eq}/\sqrt{2}$ exist indefinitely with a finite
radius or else collapse to $R=0$.

\subsection{Case 2:  $R>R_{eq}$}

The situation considered here applies to the case where a bubble of
Riemannian geometry with $R>R_{eq}$ is formed in the ambient Weyl vacuum.  The substitution of variables given by
\begin{eqnarray}
y\equiv 1-\frac{R_{eq}^{2}}{R^{2}} \label{3.15}
\end{eqnarray}
leads to the equation
\begin{eqnarray}
I_{2}\equiv \int \frac{1}{1-y}\sqrt{\frac{a^{2}-y^{2}}{y}}dy =
\mp \frac{2a}{R_{eq}}\tau . \label{3.16}
\end{eqnarray}


\subsubsection{Upper limit:  $R^{2} \gg R_{eq}^{2}$}

This limit corresponds to $y \sim 1$.  Expanding the integrand of
$I_{2}$ about this upper limit to order $(y-1)^{-1}$ one obtains
\begin{eqnarray}
I_{2}\sim \int \left[ \frac{a^{2} + 1}{2\sqrt{a^{2}-1}}-
\frac{\sqrt{a^{2}-1}}{y-1}\right] dy = \mp \frac{2a}{R_{eq}} \tau .
\label{3.16a}
\end{eqnarray}
Integration of (\ref{3.16a}) yields the result
\begin{eqnarray}
y = 1 + \exp \left[ \frac{a^{2} + 1}{2(a^{2}-1)}y \pm
\frac{2a\tau }{R_{eq}\sqrt{a^{2}-1}} \right] . \label{3.16b}
\end{eqnarray}
For sufficiently large values of $\tau $, the result with the $+$ sign
leads to values of $y>>1$ (for $a^{2}\neq1$) which invalidates the
approximation, whereas $y\rightarrow 1$ ($R \rightarrow \infty $)
exponentially as $\tau \rightarrow \infty $ for the $-$ sign in (\ref{3.16b}).


\subsubsection{Lower limit:  $R \approx R_{eq}$}

By expanding the integrand of $I_{2}$ in the neighborhood of $y=0$ to
lowest order and integrating, one obtains
\begin{eqnarray}
y^{1/2}\left( 1+\frac{y}{3}\right) \simeq \mp
\frac{\tau }{R_{eq}}. \label{3.16b1}
\end{eqnarray}
This equation has the solutions
\begin{eqnarray}
y=0 \hspace{1cm}\rm{and}\hspace{1cm}
y=-3\pm 3\sqrt{1+\left( \frac{\tau}{R_{eq}} \right) ^{2}}. \label{3.16b2}
\end{eqnarray}
The case $y=0$ is trivial: if the bubble is created at $\tau = 0$ with
$R=R_{eq}$, it will remain at the stable radius forever.  The solution with
the lower negative sign requires $y<0$ and is unphysical in the present
case ($0<y<1$).  The remaining solution leads to
\begin{eqnarray}
\frac{R^{2}_{eq}}{R^{2}}=4-3\sqrt{1+\left( \frac{\tau}{R_{eq}} \right) ^{2}}.
\label{3.16b3}
\end{eqnarray}
In this case, a bubble that has a radius $R \stackrel{>}{\sim }R_{eq}$ at
$\tau =0$ will expand to $R=\infty $ in a time $\tau = \sqrt{7/9}R_{eq}$.


\subsubsection{Special case:  $a=1$}

In this case $I_{2}$ can be integrated exactly and yields (see eq.\ (2.618.2)
in \cite{Gra})
\begin{eqnarray}
I_{2} = -2\sqrt{2}E\left( \alpha , \frac{1}{\sqrt{2}} \right) =
\mp \frac{2\tau }{R_{eq}} , \label{3.16c}
\end{eqnarray}
where $\sqrt{2}\sin \varphi /2 = R_{eq}/R$, $0<\varphi \leq \pi /2$
($0<R_{eq}/R \leq 1$) and $\alpha = \arcsin (\sqrt{2}\sin \varphi /2)$.
To second order in $\alpha $, eq.\ (\ref{3.16c}) becomes
\begin{eqnarray*}
\alpha = \mp \frac{\tau }{\sqrt{2}R_{eq}} \hspace{1cm}
\rm{or}\hspace{1cm}\frac{R}{R_{eq}}=\mp \frac{1}{\sin
\left( \frac{\tau}{\sqrt{2}R_{eq}}\right) }.
\end{eqnarray*}
In this case, $R$ diverges whenever $\tau =n\pi \sqrt{2}R_{eq}$,
and has minima of $R=R_{eq}$ for $\tau = (2n+1)\pi ^{2} /\sqrt{2}$.

In summary, bubbles that are created in the Weyl vacuum with an
initial radius greater than $R_{eq}$ appear to be unstable in all
cases, except the trivial case of $R=R_{eq}$.

\section{Discussion}

It has been suggested, in the spirit of the electroweak model, that Weyl
geometry should be considered a more fundamental geometry than Riemannian geometry
since the latter represents a state of broken scale invariance.  The
feasibility of this proposal has been explored here by considering a
Weyl-Dirac theory that supports a dynamical thin-shell solution to the
Einstein equations.  The fields that occur in the conformally-invariant
action are the source of the spherically symmetric bubbles that are
posited; the geometric fields are the fabric from which the ``particles''
are constructed.  The conformal symmetry is broken in the interior de
Sitter geometry, and the surface tension is provided by the scalar field
$\beta $.  In principle, the bubbles could be associated with microscopic
particles (see \cite{Woo93a}), perhaps born in a transition phase in the early
universe, or they could be viewed on the cosmological
scale.  In the latter case, one could consider a single bubble of
Riemannian geometry as representing our observable universe.
In either case, the extrapolation backwards from our present state of broken scale invariance to the pre-particle Weyl vacuum lies within the domain of
cosmology.  

A common feature of the most successful cosmological theories is the
presence of an inflationary epoch.  An outstanding problem in such
theories concerns the period of transition from the initial inflationary stage to the final stage of decelerating Friedmann-Robertson-Walker expansion.  This transition is often described in
terms of a first-order phase transition of the vacuum, wherein true
vacuum bubbles are created and expand, leaving isolated regions of false
vacuum.  However, the precise physical process involved in the
transition remains poorly understood.
Kodama {\it et al} \cite{Kod} provided one of the earliest
models of the transition phase in which an infinite number of true vacuum
bubbles are nucleated simultaneously on a sphere, forming a shell-like
true vacuum region around a de Sitter bubble.  They showed that the
trapped false vacuum domain either disappears leaving a black hole, or
else remains as an ever-expanding domain connected with the outer
region through a wormhole.  However, it has been shown that the 
formation of bubbles in this manner is highly improbable \cite{Gut}.
Furthermore, Linde has pointed out \cite{Lin} that the constant vacuum 
energy of de Sitter space, which has traditionally been viewed as 
necessarily leading to inflation (see, e.g., \cite{Bla}), can support other
interpretations depending upon one's choice of coordinates.
In the context of the present model, it is of interest to note that
a variety of radial motions are possible, including an exponentially
expanding solution.  However, it remains an open question as to
what interpretation should be given with regard to the identification
of ``true'' and ``false'' vacua in the present model.
As well, the precise conditions that precipitate the
symmetry breaking process requires further study.

In recent years, it has become popular to consider inflationary scenerios
within the context of conformally invariant scalar field theories.  
In this paper, a theory of this type has been used to study the dynamical
properties of a bubble once it has formed in the Weyl vacuum.
The development of the bubble model in the time-dependent case places a
significant constraint on the set of possible spherically symmetric
solutions to Einstein's equations.  This constraint strongly links the
properties of the thin shell to those of the exterior geometry.  In the
microscopic interpretation, a similar link has been shown \cite{Woo93a} to
lead to the guiding force postulated in the causal interpretation of
quantum mechanics.  In the cosmological interpretation, such a link is of
interest because it allows for the possibility that information about the
external metric may be available to an internal observer.  This could prove
to be of significance in the problem of  boundary
conditions of the universe.

The analysis of the radial equation of motion of the thin shell revealed
the following results.  Bubbles that emerge in the Weyl vacuum with
$R>R_{eq}$ are unstable (except the trivial static case of $R=R_{eq}$) as
they variously diverge to $R=\infty $.  For bubbles with $R<R_{eq}$
initially, a number of possiblities exist.  Of particular interest are
those oscillatory solutions for which the radius remains finite.  The
freedom allowed by the parameters in the theory permit the scale associated
with these bubbles to take on any value from the microscopic to the
cosmological.  In the first case, the results obtained here may prove to be
of interest in the geometric causal interpretation of Ref.\ \cite{Woo93a}.
In the latter case, one may find the present model useful in addressing some
of the unresolved issues in inflationary cosmology.  Indeed, the intimate
relationship between particle physics and cosmology suggests that we
should be considering theories of the very early universe that find
application at any scale.

\vspace{.3cm}
\begin{flushleft}
{\large
{\bf Acknowledgements}}\vspace{.3cm}\\
\end{flushleft}
This work was supported in part by the Natural Sciences and
Engineering Research Council of Canada.  One of the authors (G.P.)
wishes to thank Dr. K. Denford, Dean of Science, University of Regina,
for continued research support.

\appendix
\section{The time-dependent model}

A time-dependent, spherically symmetric solution to the Einstein field
equations (\ref{2.3}) is, in general, obtained by solving the three
equations
\begin{eqnarray}
\frac{d}{dr}(re^{-\mu })&=&1+r^{2}
\left[ -\frac{q^{2}}{4\beta ^{2}r^{4}}-e^{-\mu }
\left( 2\frac{\beta ''}{\beta }-
\frac{\beta '^{2}}{\beta ^{2}}-\mu '\frac{\beta '}{\beta }+
\frac{4}{r}\frac{\beta '}{\beta }\right) \right. \nonumber \\
\nonumber \\
&&\hspace{3.2cm}\left. +\hspace{1pt}e^{-\nu }\left(
\mu _{,t}\frac{\beta _{,t}}{\beta }+
3\frac{\beta _{,t}^{\ 2}}{\beta ^{2}}\right) +
\frac{1}{2}\lambda \beta ^{2}\right] , \label{A.1}
\end{eqnarray}
\begin{eqnarray}
\frac{d}{dr}(\mu +\nu )&=&r\left[
2\frac{\beta ''}{\beta }-4\frac{\beta '^{2}}{\beta ^{2}}-
(\mu '+\nu ')\frac{\beta '}{\beta }\right] \nonumber \\
\nonumber \\
&&\hspace{.6cm}+\hspace{1pt}
re^{\mu -\nu }\left[ 2\frac{\beta _{,tt}}{\beta }-
4\frac{\beta _{,t}^{\ 2}}{\beta ^{2}}-
(\mu +\nu )_{,t}\frac{\beta _{,t}}{\beta }\right], \label{A.2}
\end{eqnarray}
and
\begin{eqnarray}
\frac{d}{dt}(e^{-\mu })=-re^{-\mu }\left(
2\frac{\beta '_{,t}}{\beta }-\nu '\frac{\beta _{,t}}{\beta }-
\mu _{,t}\frac{\beta '}{\beta }-4\frac{\beta '}{\beta }
\frac{\beta _{,t}}{\beta }\right) . \label{A.3}
\end{eqnarray}
The solutions to these equations in the interior and exterior spaces are
then generally used in determining the equation of motion $r = R(\tau )$
for the thin shell in the frame comoving with the thin shell with proper
time
$\tau $.
In the GMC formalism, one introduces the spherically symmetric intrinsic metric
\begin{eqnarray}
ds_{\Sigma }^{2}=-d\tau ^{2}+R^{2}(\tau )(d\theta ^{2}+
\sin ^{2}\theta d\phi ^{2}) \label{A.4}
\end{eqnarray}
and the extrinsic curvature \cite{Woo92a}
\begin{eqnarray}
K_{\mu \nu } = -\frac{1}{2}n^{\alpha }h_{\mu \nu ,\alpha }, \label{A.5}
\end{eqnarray}
where $n^{\mu }$ is the unit spacelike vector field normal to $\Sigma $
and $h_{\mu \nu }$ are the coefficients of the intrinsic metric on $\Sigma $.
The equation of motion of the thin shell can be obtained from the jump in
the $\theta \theta $-component of the extrinsic curvature \cite{Woo92a}:
\begin{eqnarray}
(K^{\theta }_{\ \theta })^{E}-(K^{\theta }_{\ \theta })^{I} = \sigma ,
\label{A.6}
\end{eqnarray}
where $2\sigma $ is the surface energy density of the thin shell,
\begin{eqnarray}
(K^{\theta }_{\ \theta })^{E,I} = -\frac{1}{R}e^{-\mu _{E,I}}X_{E,I},
\label{A.7}
\end{eqnarray}
and
\begin{eqnarray}
X_{E,I}^{2} = e^{(\mu -\nu )_{E,I}}\dot{R}^{2}+e^{-\nu _{E,I}}. \label{A.8}
\end{eqnarray}
The surface energy $\sigma $ is generally taken to be an arbitrary
parameter in models involving domain walls.  However, by identifying this
surface energy with the discontinuity in the normal derivative of the field
$\beta $ across $\Sigma $, the ``particle'' proposed in Ref.\ \cite{Woo92a}
was shown to be an entity that is derived entirely from the fields in the
theory.

In the time-dependent, spherically symmetric theory under consideration,
the scalar field $\beta $ establishes an intimate link between $V^{I}$, the
thin shell, and $V^{E}$.  The dynamical implications of this link can be
seen
as follows.  In the time-dependent case,
\begin{eqnarray}
n^{\mu } = e^{(\mu +\nu )/2}(e^{-\nu }\dot{R}, e^{-\mu }X,0,0) \label{A.9}
\end{eqnarray}
and
\begin {eqnarray}
\sigma = e^{-(\mu +\nu )_{E}/2}\frac{\omega }{X_{E}}, \label{A.10}
\end{eqnarray}
where $\omega \equiv (\beta '/\beta )|_{r=R}$.  Eq.\ (\ref{A.10}) provides
a correction to (4.3) in \cite{Woo93b} where $n^{\mu }$ was not normalized
to unity.  Using (\ref{A.9}) and (\ref{A.10}), the equation of motion
(\ref{A.6}) for the thin shell becomes
\begin{eqnarray}
e^{(\nu -\mu )_{I}/2}X_{I} = e^{(\nu -\mu )_{E}/2}X_{E} +
e^{-(\mu +\nu )_{E}/2}\frac{\omega R}{X_{E}}. \label{A.11}
\end{eqnarray}
For $\beta (t,r) = f(t)/r$, it follows that $\omega R = -1$.
Using this, together with the interior metric (\ref{2.7}), the square of
(\ref{A.11}) becomes
\begin{eqnarray}
1+\dot{R}^{2}+\frac{1}{6}\lambda \beta _{0}^{2}R^{2} =
e^{(\mu - \nu )_{E}}\frac{\dot{R}^{4}}{X_{E}^{2}}. \label{A.12}
\end{eqnarray}
Now, there are two distinct expressions for $X_{E}^{2}$ in the present model.
First, there is (\ref{A.8}) that follows from the form of $dt/d\tau $ in
the spherically symmetric case.  Then there is (\ref{2.10}) that follows
from the choice of boundary conditions for $\beta $.  In general, there is
no guarantee that the two equations of motion that result when these two
expressions for $X_{E}^{2}$ are substituted into (\ref{A.12}) will be
consistent with each other.  Of course, the form of the equation of motion
also depends on the form of the solutions for $e^{\mu _{E}}$ and $e^{\nu
_{E}}$.  Clearly then, this model is rather tightly constrained.  Indeed,
the equation of motion of the thin shell could be viewed as being somewhat
prescriptive of the nature of the exterior geometry.  In other words, the
intimate link between the thin shell and the exterior geometry suggests
that the properties of the thin shell may provide information regarding
properties of the exterior geometry.  This in fact is the case.  Equating
the two expressions for $X_{E}^{2}$ given in (\ref{A.8}) and (\ref{2.10})
one finds
\begin{eqnarray}
e^{(\mu - \nu )_{E}}\dot{R}^{2}+e^{-\nu _{E}} =
\frac{\beta '^{2}}{\beta _{,t}^{2}}\dot{R}^{2}.  \label{A.13}
\end{eqnarray}
Substitution of $e^{(\mu - \nu )_{E}}\dot{R}^{2}$ from (\ref{A.13}) into
the equation of motion (\ref{A.12}) leads to the result
\begin{eqnarray}
e^{\nu _{E}} = -\frac{\beta _{,t}^{2}}
{\beta '^{2}(1+\frac{1}{6}\lambda \beta _{0}^{2}R^{2})}. \label{A.14}
\end{eqnarray}
Hence, the equation of motion, which applies only at $r=R$, leads to an
algebraic expression for $e^{\nu _{E}}|_{r=R}$.   One may view (\ref{A.14})
as a necessary condition for the projection of $e^{\nu _{E}}(r)$ onto
$\Sigma $ in order for the present model to be internally consisent.
Adopting this viewpoint, the most obvious form for $e^{\nu _{E}}(r)$ to
satisfy this condition is
\begin{eqnarray}
e^{\nu _{E}}=-\frac{\beta _{,t}^{2}}{\beta '^{2}(1+
\frac{1}{6}\lambda \beta ^{2}r^{2})}. \label{A.15}
\end{eqnarray}
For this choice, the Einstein equations (\ref{A.1})--(\ref{A.3}) are
satisfied when
\begin{eqnarray}
e^{\mu _{E}} = -\frac{1+\frac{1}{6}\lambda \beta ^{2}r^{2}}{\gamma ^{2}
\beta ^{2}r^{4}}, \label{A.16}
\end{eqnarray}
where $\gamma $ is an arbitrary constant with dimensions (length)$^{-1}$.
In the present model of a spherically symmetric region of Riemannian space
in an exterior Weyl geometry, the equations of the dynamical model will be
self-consistent provided the exterior metric is of the form\footnote{The
time-dependent solution presented in \cite{Woo93b} satisfies this condition
for $c_{1} = -c_{2}^{2}$ and $q=0$.  However, the oscillatory nature of the
thin shell solution given in that paper is superceded by the results
contained herein.} given in (\ref{A.15}) and (\ref{A.16}).

\end{document}